\documentclass{optica-article}

\journal{opticajournal} 

\articletype{Research Article}
\usepackage{xcolor}
\usepackage{lineno}
\usepackage{bm}

\begin{document}

	\title{Broadband, robust, and tunable beam splitter based on topological unidirectional surface magnetoplasmons}

\author{Lujun Hong,\authormark{1,2,$\dag$, *} Chao Liu,\authormark{1,2,$\dag$} Jun Wu,\authormark{3} Chaojian He,\authormark{4} Kai Yuan,\authormark{1}  Xiaohua Deng,\authormark{2} Song Yang,\authormark{4} Zhen Gao \authormark{5,*}}

\address{\authormark{1}School of Information Engineering, Nanchang University, Nanchang 330031, China\\
\authormark{2}Institute of Space Science and Technology, Nanchang University, Nanchang 330031, China\\
\authormark{3}School of Physics and Materials Science, Nanchang University, Nanchang 330031, China\\
\authormark{4}Laboratory of All-Solid-State Light Sources, Institute of Semiconductors, Chinese Academy of Sciences, Beijing 100083, China\\
\authormark{5}Department of Electronic and Electrical Engineering, Southern University of Science and Technology, Shenzhen, China\\
\authormark{$\dag$} The authors contributed equally to this work.
}
\email{\authormark{*}ljhong@ncu.edu.cn; gaoz@sustech.edu.cn} 


\begin{abstract} 
Beam splitters are pivotal components in integrated microwave and photonic systems. However, conventional designs based on directional coupling or multi-mode interference often suffer from backscattering, frequency-dependent splitting ratios, and limited bandwidth. To overcome these limitations, here, we propose a new physical mechanism to achieve a broadband, robust, and tunable beam splitter by manipulating the mode coupling of the topological unidirectional surface magnetoplasmons (USMP) at the input and output waveguides. We show that the beam splitter not only exhibits strong robustness against obstacles but also achieves a broad bandwidth across nearly the entire USMP band with arbitrarily tunable and frequency-independent splitting ratios. Moreover, the operating band of the beam splitter can be actively tuned by adjusting the external magnetic field, while its robust and broadband characteristics are retained. Our results extend the research frontier of beam splitters and may have potential applications in integrated photonic devices and modern communication systems.
\end{abstract}

\section{INTRODUCTION}
Beam splitters are critical components in integrated photonic circuits and are widely used for on-chip information processing, quantum computing, and reconfigurable networks\cite{wangOnchipTopologicalBeamformer2024,hu_Onchip_2021}. Compared to common beam splitters with equal splitting ratios, it is challenging to realize beam splitters with arbitrarily tunable splitting ratios \cite{wu_Broadband_2023,chungRobustSiliconArbitrary2020,maoAdiabaticCouplerDesignIntended2019}, which usually require complex designs to achieve precise control of the splitting ratios. To date, various approaches have been employed to realize tunable beam splitters, such as multimode interference (MMI) and directional coupling\cite{papadovasilakisFabricationTolerantWavelength2022,Lu:15}.  However, traditional tunable beam splitters based on these methods often suffer from backscattering caused by structural imperfections and mode mismatch at the input and output ports\cite{boscolo_Junctions_2002}, significantly affecting splitting performance and transmission efficiency.

Topological photonics, a research field that combines topological physics and photonics, offers an attractive way to overcome the challenge of backscattering\cite{luTopologicalPhotonics2014b,ozawaTopologicalPhotonics2019,shalaevRobustTopologicallyProtected2019,temporary-citekey-3841,liuTopologicalChernVectors2022b}, and one of its most remarkable features is the topological unidirectional modes with time-reversal symmetry breaking \cite{qu_Topological_2024,chen_OneWay_2025,goudarzi_Tolerance_2024,wang_Threedimensional_2025,shi_Coexistence_2024}.  
Such unidirectional modes were first theoretically proposed as a photonic analogue of quantum Hall edge states in condensed matter systems \cite{haldanePossibleRealizationDirectional2008,wang_ReflectionFree_2008}, and experimentally verified in gyromagnetic photonic crystals (PhCs) with backscattering immunity \cite{ wangObservationUnidirectionalBackscatteringimmune2009}. As an alternative unidirectional photonic mode, surface magnetoplasmons (SMPs) have recently attracted great interest because of their simple structure, unique topological property, and ability to manipulate electromagnetic waves at subwavelength scales, which has significant applications in robust and integrated photonic devices \cite{jinTopologicalMagnetoplasmon2016,gao_Photonic_2016,tsakmakidisBreakingLorentzReciprocity2017,monticoneTrulyOnewayLane2020a,hong_Magnetic_2019,liangTunableUnidirectionalSurface2021,xuAllOpticalDigitalLogic2023a}. Unidirectional SMP (USMP) have been revealed to be topologically protected and robust against defects and obstacles due to their nonzero gap Chern number ($C_{gap}\neq 0$)\cite{silveirinha_Chern_2015,silveirinhaProofBulkEdgeCorrespondence2019,fernandesExperimentalVerificationIlldefined2022}. Recently, the existence of USMP has been experimentally demonstrated in a gyromagnetic YIG-air-metal waveguide at microwave frequency \cite{liUnidirectionalGuidedwavedrivenMetasurfaces2024}. Using YIG-based unidirectional photonic modes, various nonreciprocal topological photonic devices have been theoretically proposed and experimentally demonstrated, including topological lasers \cite{bahariNonreciprocalLasingTopological2017}, large area waveguides \cite{chenPredictionObservationRobust2022,wang_Topological_2021,yu_Topological_2023}, rainbow trapping\cite{chen_Switchable_2019}, and broadband slow light \cite{chen_Multiple_2024,mann_Broadband_2021,liu_Selective_2025}. More recently, magnetically tunable beam splitters based on MMI have been theoretically proposed based on topological USMP \cite{liuRobustMultimodeInterference2025,35r5-ngx2} and experimentally realized in a gyromagnetic PhC \cite{tangMagneticallyControllableMultimode2024}. However, while the performance of the magnetically tunable beam splitters is improved, their splitting ratios are highly sensitive to the operating frequency due to the mechanism of mode interference, seriously limiting its operation bandwidth and making it challenging to realize a broadband and tunable beam splitter with arbitrary splitting ratios. 

Here, by manipulating the mode coupling between the input and output modes at the junction of the beam splitter, we propose a broadband and tunable beam splitter based on topological USMP. We show that arbitrarily tunable and frequency-independent splitting ratios can be achieved across nearly the entire USMP band, overcoming the bandwidth limitations of the existing MMI-based beam splitters. Furthermore, the working band of the beam splitter can be dynamically tuned by adjusting the external magnetic field (EMF).  Notably, the beam splitter exhibits robust and broadband splitting performance against obstacles, achieving nearly 100\% transmission efficiency without backscattering.

\section{BASIC PHYSICAL MODEL OF TOPOLOGICAL USMP }
\begin{figure}[b]
	\centering\includegraphics[width=4.2 in]{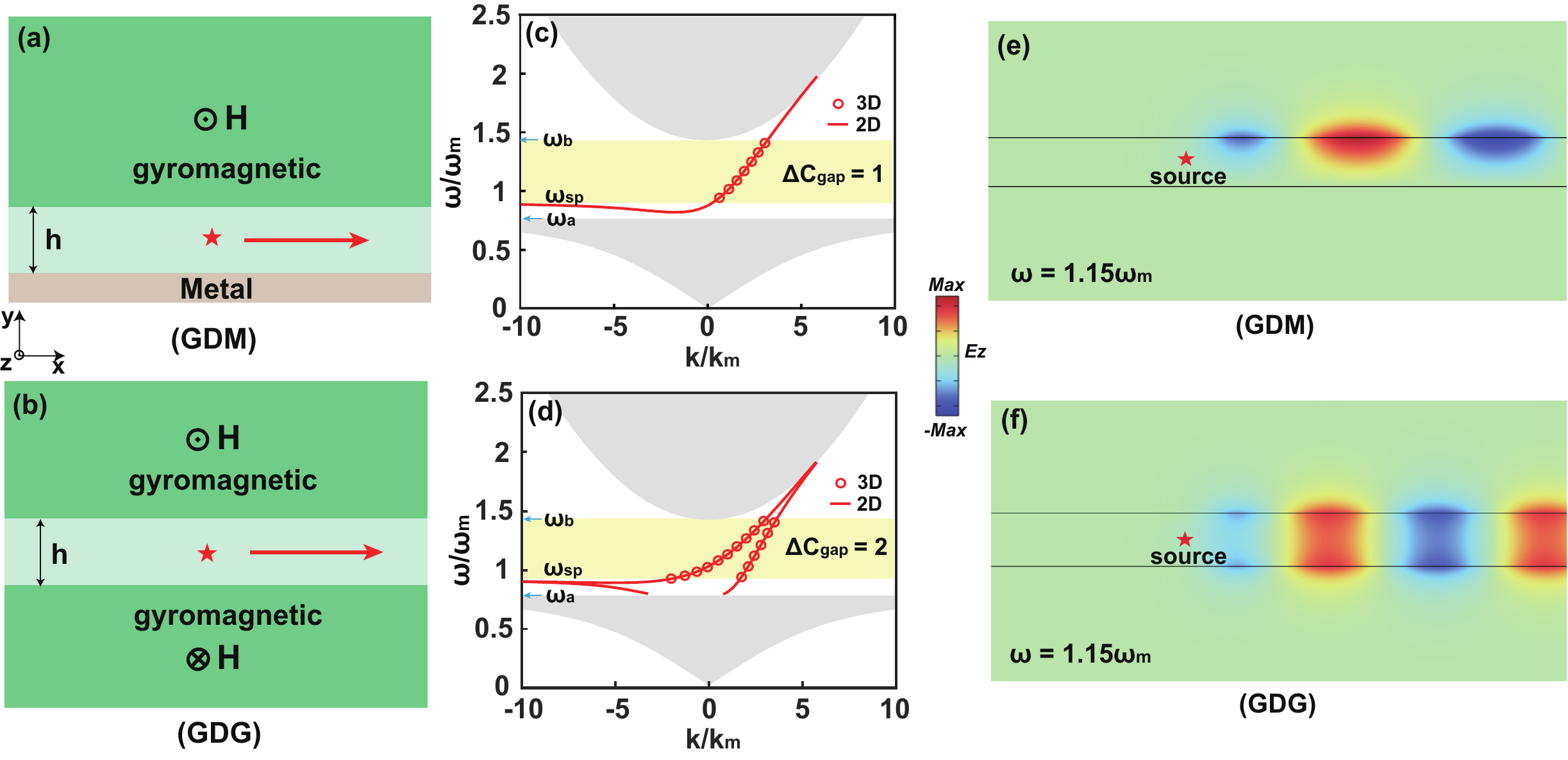}%
	\caption{\label{fig1} Schematic of (a) the GDM waveguide: gyromagnetic-dielectric-metal under one EMF, and (b) the GDG waveguide: gyromagnetic–dielectric–gyromagnetic under two opposing EMFs. (c), (d) Dispersion diagrams of the GDM and GDG waveguides, respectively. The circles and solid lines represent the SMP modes in 3D and 2D waveguides, while the gray and yellow areas represent the bulk modes and the USMP band, respectively. (e), (f) Simulated $Ez$-field distribution in the GDM and GDG waveguides for an excitation source (pentagram) at $\omega = 1.15\,\omega_m$. The parameters are \(h=6\) mm and $\omega_0 = 0.4\omega_m$, $H=715$ Oe.}
\end{figure}
The fundamental physical model of the proposed beam splitter is composed of gyromagnetic-dielectric-metal (GDM) waveguide and gyromagnetic-dielectric-gyromagnetic (GDG) waveguide, as illustrated in Fig.~\ref{fig1}(a) and (b). 
In both waveguides, YIG acts as the gyromagnetic material, with a relative permittivity of $\varepsilon_m = 15$. Air serves as the dielectric spacer, has a thickness of \( h\). The YIG layer in the GDM waveguide is magnetized by one EMF, while it is magnetized by two opposing EMFs for the GDG waveguide. Thus, the YIG 
exhibits gyromagnetic anisotropy, with a permeability tensor ${\mu_m}$ \cite{tangMagneticallyControllableMultimode2024,xiTopologicalAntichiralSurface2023,gao2025science}:
\begin{equation}
	\small
	\mu^{\pm}_m = 
	\begin{bmatrix}
		\mu_1 & \mp i\mu_2 & 0 \\
		\pm i\mu_2 & \mu_1 & 0 \\
		0 & 0 & 1
	\end{bmatrix}
\end{equation}
with the symbols $\pm$ representing magnetic fields aligned  along the +z and -z directions, $\mu_1$= $1+\dfrac{\omega_m( \omega_0 + i\alpha \omega)}{(\omega_0+i\alpha\omega)^2 - \omega^2}$, $\mu_2$= $\dfrac{\omega_m \omega}{(\omega_0+i\alpha\omega)^2 - \omega^2}$, where $\alpha$ is the damping coefficient, $\omega_0$ = $2\pi \gamma H$ (${\gamma}=2.8$~MHz/Oe) is the resonance frequency, ${\omega}$ is the angular frequency, and $\omega_m$ is the characteristic circular frequency, respectively. In the magnetized YIG, the bulk modes for TE polarization have a dispersion relation of $k^2 \leq \mu_{\rm{v}}\epsilon_{m}k^2_{0}$, where $\mu_{\rm{v}}=\mu_1-\mu_{2}^2/\mu_1$, $k_0=\omega/c$, and $k$ is the propagation constant. A bandgap [$\omega_{a}$,$\omega_{b}$] is opened by the EMF, where $\omega_a = \sqrt{\omega_0(\omega_0+\omega_m)}$ and $\omega_b = \omega_0 + \omega_m$. Such a bandgap is topologically nontrivial owing to the nonzero gap Chern number $C_{gap} = \pm1$ in the YIG bulk modes \cite{fernandesExperimentalVerificationIlldefined2022,silveirinha_Chern_2015}. Since the metal has \(C_{gap} = 0\), the GDM waveguide yields a gap Chern number difference of $\Delta C_{gap} = 1$.  Unlike the GDM waveguide, the opposite magnetic fields result in a larger \(\Delta C_{gap}= 2\) for the GDG waveguide.  According to the bulk-edge correspondence principle that the number of unidirectional edge modes is equal to $\Delta C_{gap}$ \cite{silveirinhaProofBulkEdgeCorrespondence2019,tauber_Anomalous_2020,han_Bulk}, the GDM and GDG waveguides are expected to support one and two unidirectional SMP modes in the nontrivial topological bandgap, respectively. To illustrate this, the dispersion equation of SMP in both waveguides can be analytically derived as follows (for details see Appendix)
\begin{subequations}
	\begin{eqnarray}
		k\dfrac{\mu_2}{\mu_1} +\alpha_m+
		\alpha _{r}\mu_{\rm{v}}{\rm{coth}}\left ({\alpha _{r}h} \right )=0 \quad(GDM)
		\label{eq:eq3a}\\
		e^{2\alpha_r h} - \left( \dfrac{\alpha_r \mu_v - k\dfrac{\mu_2}{\mu_1} - \alpha_m}{\alpha_r \mu_v + k\dfrac{\mu_2}{\mu_1} + \alpha_m} \right)^2 = 0 \quad(GDG)
		\label{eq:eq3b}
	\end{eqnarray}
	\label{eq:eq3}
\end{subequations} 
where $\alpha_{r}=\sqrt{k^{2}-k_{0}^{2}}$ and  $\alpha_{m}=\sqrt{k^{2} -\mu_{\rm{v}}\epsilon_{m}k_{0}^{2}}$ for the air and YIG slab layers, respectively \cite{liUnidirectionalGuidedwavedrivenMetasurfaces2024}. For both GDM and GDG waveguides, the SMP modes have the same asymptotic frequencies of  $\omega_{sp} = \omega_0 + 0.5\omega_m$  as $k \rightarrow -\infty$.

Using Eq.~(\ref{eq:eq3}), we numerically solve the SMP dispersion and bulk band structure for the two waveguides. In the calculation, $\omega_m = 10\pi \times 10^9~\mathrm{rad/s}$ ($f_m = 5~\mathrm{GHz}$), $\alpha=0 $, and $h = 6~\mathrm{mm}$. Figure~\ref{fig1}(c) and (d) show the dispersion diagram of SMP and bulk modes in the GDM and GDG waveguides for $\omega_0 = 0.4\omega_m$, corresponding to $H=715$~Oe. Obviously, there are the same nontrivial topological bandgap [$\omega_{a}$,$\omega_{b}$] for both GDM (\(\Delta C_{gap}=1\)) and GDG waveguides (\(\Delta C_{gap}=2\)), where $\omega_{a}=0.75\omega_{m}$ and $\omega_{b}=1.4\omega_{m}$. In the frequency range [$\omega_{sp}$,$\omega_{b}$], the SMP modes can propagate forward only due to a positive group velocity ($d\omega/dk>0$), demonstrating unidirectional SMP (USMP) behavior. The corresponding USMP band is [0.9$\omega_{m}$,1.4$\omega_{m}$], as highlighted in yellow. As expected, the GDM waveguide supports only one USMP mode within the nontrivial bandgap, while the GDG waveguide supports two USMP modes. These results are consistent with the prediction of the bulk-edge correspondence. To clearly illustrate this, we simulate the wave propagation in the GDM and GDG waveguides, as shown in Fig.~\ref{fig1}(e) and~\ref{fig1}(f). A line current source (pentagram) with $\omega=1.15\omega_{m}$ is placed at the center of the air layer in both waveguides to excite the USMP mode. The \(E_z\) field distributions clearly demonstrate unidirectional propagation without backscattering. Moreover, it should be emphasized that such a 2D model is equivalent to a 3D realistic model, terminated by metal plates in the \(z\)-direction. To verify this, we analyzed 3D GDM and GDG waveguides with width $d$, and performed modal analysis to solve the eigenmode solutions using COMSOL Multiphysics, and the results for $d = 6$~mm are displayed as the open circles in Fig.~\ref{fig1}(c) and (d). The dispersion for the 3D waveguides agrees well with those of 2D system in the whole USMP band. 

\section{BROADBAND TUNABLE SPLITTER}
\begin{figure}[b]
	\centering\includegraphics[width=3.4 in]{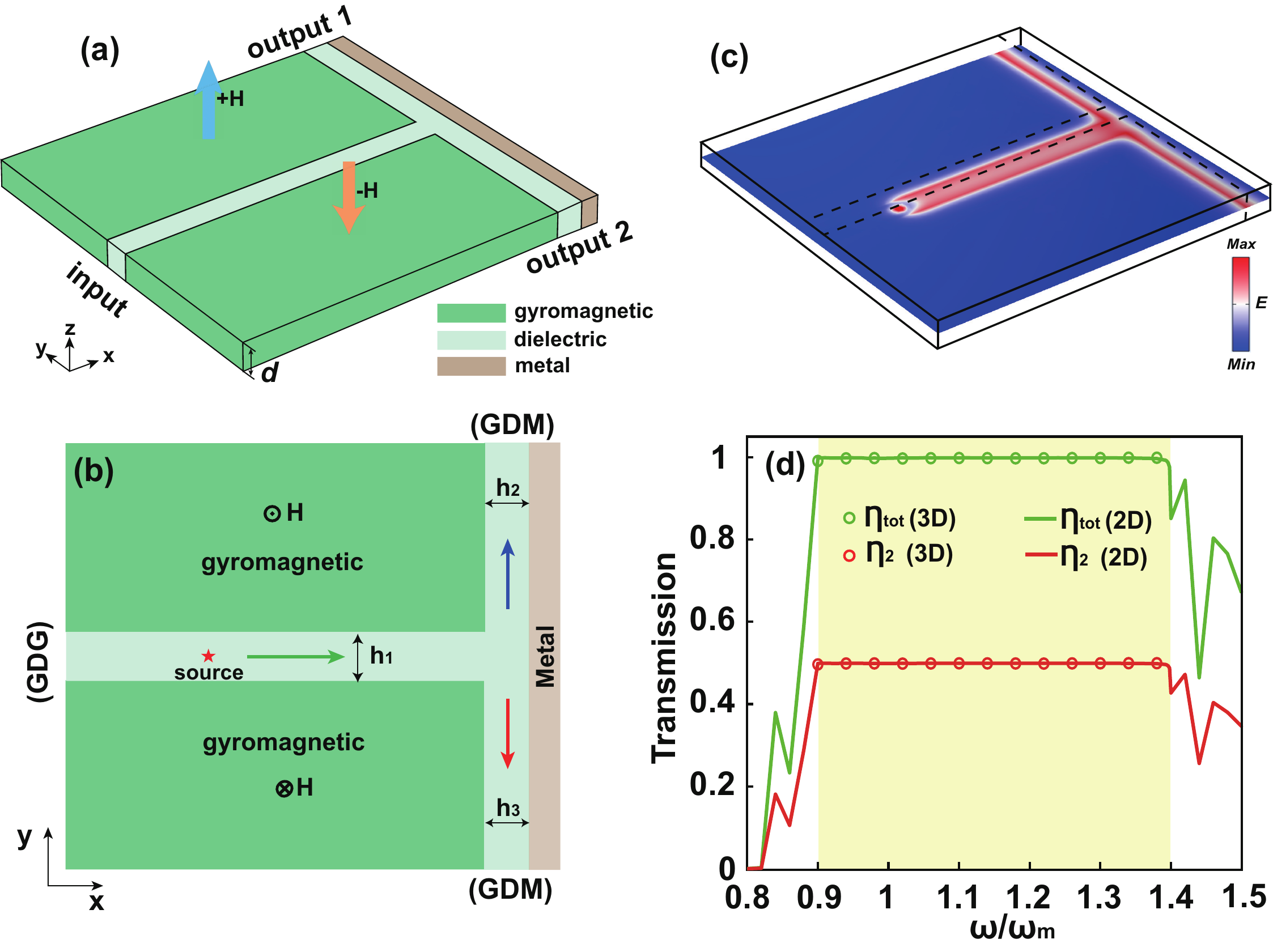}
	\caption{\label{fig2} Schematic of (a) 3D with width \(d\) and (b) 2D splitters. The input port is a GDG waveguide, and the two output ports are GDM waveguides, with the dielectric layer widths \(h_1, h_2, h_3\), respectively. (c) Simulated $E$-field distribution of the 3D splitter at $\omega = 1.15\,\omega_m$. (d) Transmission spectra ($\eta_{2}, \eta_{tot}$) of the 3D (circles) and 2D (lines) symmetric splitter. The yellow-shaded region denotes the USMP band. The parameters are \(h_1=h_2=h_3= 6\) mm, \(d= 6\) mm , and $H=715$ Oe. }
\end{figure}
Based on the two topological USMP waveguides, we design a 3D realistic broadband tunable beam splitter with a waveguide width \(d\), as illustrated in Fig.~\ref{fig2}(a). The corresponding 2D cross-section is shown in Fig.~\ref{fig2}(b). The input section is a GDG waveguide, and the two output sections are the GDM waveguides, where the dielectric layer thicknesses are \(h_1\), \(h_2\), and \(h_3\), respectively. To evaluate the power-splitting behavior, we perform full-wave simulations for the splitter. A line current source is placed at the center of the air layer to excite the SMP mode. The transmission coefficients at the output 1 and output 2 are denoted as $\eta_{1}$ and $\eta_{2}$, respectively. Figure~\ref{fig2}(d) illustrates $\eta_{1}$, $\eta_{2}$ and the total transmission efficiency $\eta_{tot}=\eta_{1}+\eta_{2}$ for the 3D and 2D symmetric splitter. Here, \( h_1 = h_2 = h_3 = 6\ \mathrm{mm}\), $\alpha=3\times 10^{-5}$, and the other parameters are the same as in Fig.~\ref{fig1}. 
It is seen that $\eta_{tot}$ is nearly 100\% in the USMP band because unidirectional transport overcomes the backscattering issue in the conventional bidirectional waveguides. Conversely, outside the USMP band, $\eta_{tot}$ drops significantly. Moreover, due to the structural symmetry, $\eta_{2}$ remains 50\% across the whole USMP band, demonstrating the broadband equal-power splitting property of the beam splitter. To clearly illustrate this, the full-wave simulation result for the 3D symmetric splitter using $\omega=1.15\omega_m$ as an example is displayed in Fig.~\ref{fig2}(c). As expected, the excited wave can only propagate forward and is equally divided into two output waveguides. Notably, the transmission coefficients of both 3D and 2D models are in agreement in the USMP band as shown in Fig.~\ref{fig2}(d), which are consistent with the dispersion results presented in Fig.~\ref{fig1}. Hence, the subsequent simulations are performed using a simplified 2D model and we focus on the USMP band.

\begin{figure}[t]
	\centering\includegraphics[width=4.2 in]{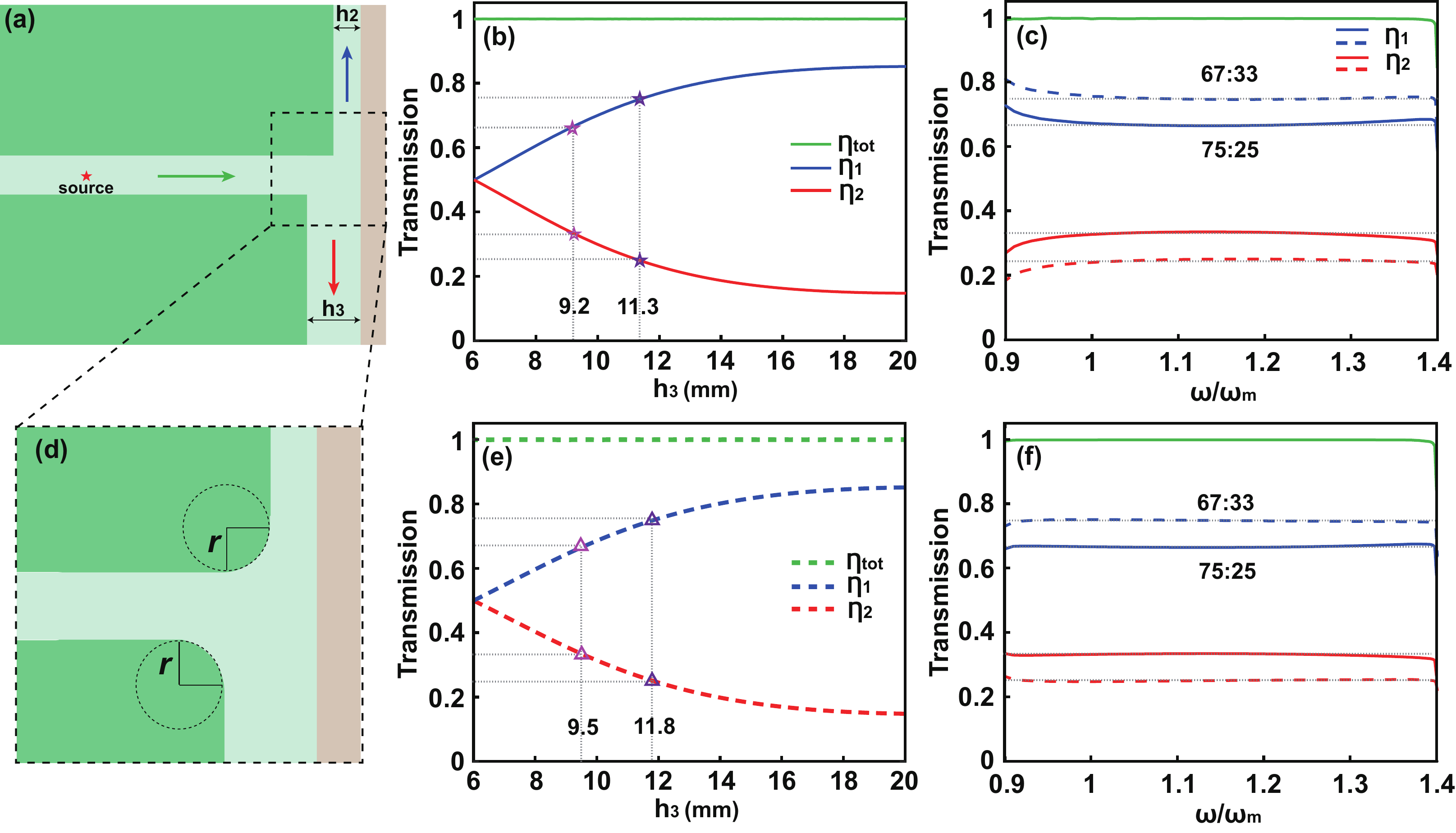}
	\caption{\label{fig3}  Broadband tunable asymmetric splitter. Schematic of (a) unfilleted (\(r = 0\ \mathrm{mm}\)) and (d) filleted (\(r = 3\ \mathrm{mm}\)) asymmetric splitter  for \( h_2 \neq h_3 \).  Transmission spectra ($\eta_{1}, \eta_{2}, \eta_{tot}$) versus \(h_3\) for (b) unfilleted and (e) filleted splitters at \( \omega = 1.15\omega_m \). Transmission spectra for $\eta_{1}:\eta_{2}=$ 67:33 (dashed lines) and 75:25 (solid lines) in the USMP band.  (c) unfilleted splitter for $h_3 = $ 9.2 mm and $h_3 =$ 11.3 mm, and (f) filleted splitter for $h_3 = $ 9.5 mm and $h_3 = $ 11.8 mm. The other parameters are the same as in Fig.~\ref{fig2}.}
\end{figure}
Compared to symmetric structures enabling equal splitting, unequal splitting usually requires breaking structural symmetry and complex design. For the proposed asymmetric splitter ($h_2\neq h_3$), broadband tunable splitting can be realized by simply adjusting the air layer thickness in the output waveguides, as shown in Fig.~\ref{fig3}(a). To verify the tunability of the splitter, the corresponding transmission as a function of \(h_3\) at $\omega = 1.15\,\omega_m$ is presented in Fig. \ref{fig3}(b), with the other parameters identical to those in the symmetric splitter shown in Fig.~\ref{fig2} (b). As \(h_3\) increases from 6 mm to 20 mm, $\eta_{2}$ decreases from 50\% to 15\% while $\eta_{1}$ increases from 50\% to 85\%. These results indicate that the splitting ratio can be continuously tuned from 15\% to 85\% at $\omega = 1.15\,\omega_m$ by varying the output air layer thickness. Similarly to the symmetric splitter, $\eta_{tot}$ remains nearly 100\% for the asymmetric splitter. To demonstrate the broadband performance, two typical unequal splittings are analyzed: $\eta_{1}:\eta_{2}=67:33$ for $h_3=9.2$ mm, and $\eta_{1}:\eta_{2}=75:25$ for $h_3=11.3$ mm, marked by pentagrams in Fig.~\ref{fig2} (b). The corresponding transmission spectra in Fig.~\ref{fig3} (c) show broad bandwidths within a 1\% deviation for a desired splitting ratio. For the 67:33 and 75:25 cases, the bands span [$0.97\omega_m, 1.34\omega_m$] and [$1.0\omega_m, 1.39\omega_m$], covering 74\% and 78\% of the USMP bandwidth, respectively. These results demonstrate that a broadband tunable asymmetric splitter can be achieved by simply adjusting the output air layer thickness.

To further improve the bandwidth of the asymmetric splitter, the junction corners with a radius of \(r\) are filleted for smoother mode coupling, as shown in Fig.~\ref{fig3}(d). Figure~\ref{fig3}(e) shows the updated transmission spectra versus \(h_3\) for \(r=3\)~mm. Similarly to the unfilleted splitter, the splitting ratio can be continuously tuned for the asymmetric splitter. It is found that $\eta_{1}:\eta_{2}=67:33$ for $h_3=9.5$ mm, and $\eta_{1}:\eta_{2}=75:25$ for $h_3=11.8$ mm, marked by triangles. Figure~\ref{fig3}(f) shows the transmission spectra for these two cases. With the filleted design, the bandwidths for both cases increase to 96\% of the USMP band, with corresponding frequency ranges of [$0.9\omega_m, 1.38\omega_m$] and [$0.91\omega_m, 1.39\omega_m$], while maintaining a splitting deviation within 1\% of the desired ratio. Compared to the splitter without fillet in Fig.~\ref{fig3}(c), the bandwidth of the asymmetric splitter with fillet is significantly expanded to nearly the entire USMP band with extremely low deviation.

\begin{figure}[b]
	\centering\includegraphics[width=4.2 in]{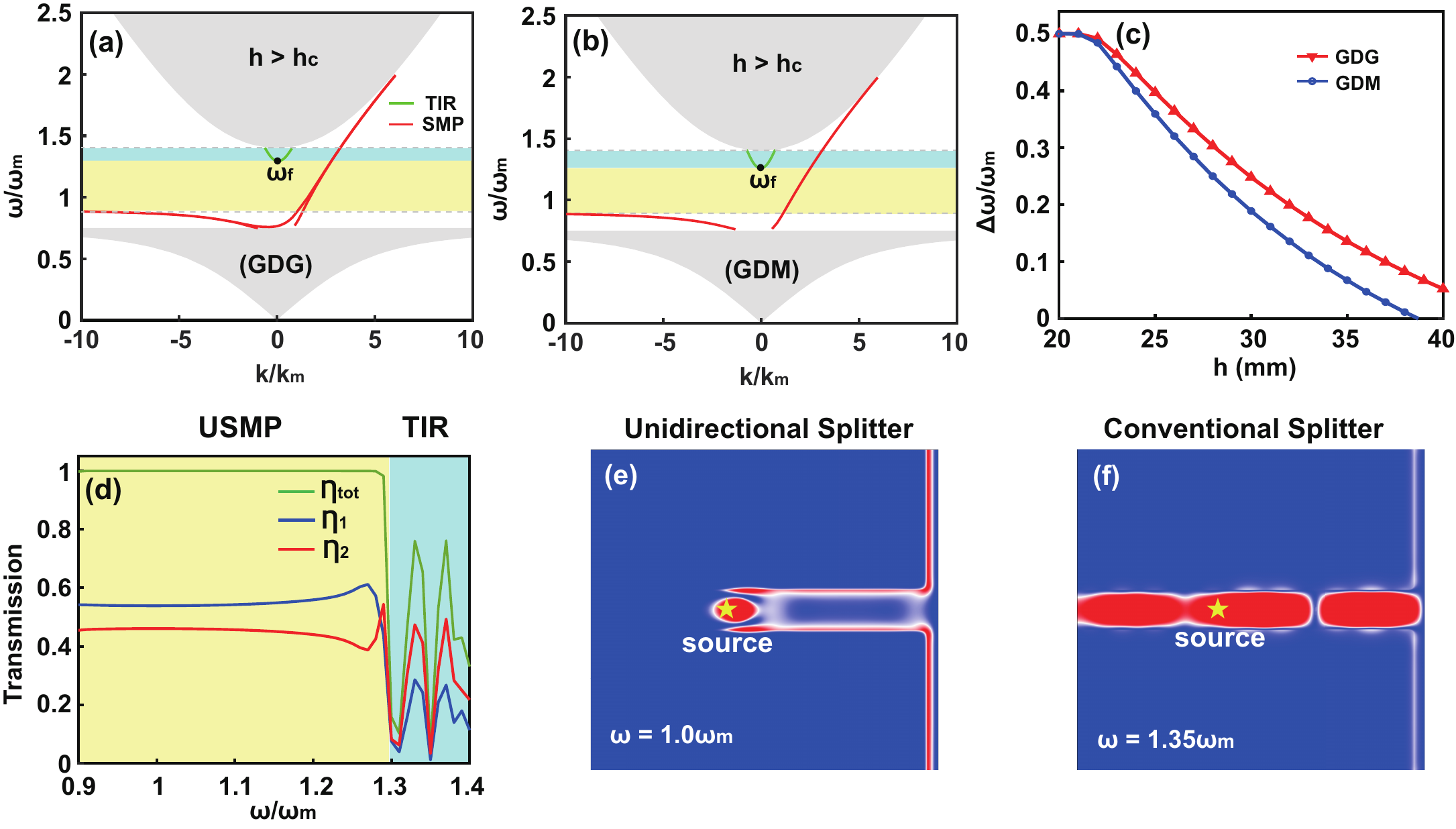}
	\caption{\label{fig4}Dispersion diagrams of TIR (Green curves) and SMP modes (red curves)  for $h>h_c$. (a) GDG and (b) GDM waveguides for \( h=25\,\text{mm} \).  The yellow and blue areas represent USMP and bidirectional TIR, where \( \omega_{f} \) indicates the cutoff frequency of TIR modes. (c) USMP bandwidth $\Delta\omega$ versus $h$. (d) Transmission spectrum of the splitter with \( h_1=25\,\text{mm} \), \( h_2=12\,\text{mm} \), and \( h_3=15\,\text{mm} \), where the yellow and blue regions correspond to frequency ranges marked in (a), respectively. (e,f) Comparison of the \(E\)-field distributions in the unidirectional splitter (e) with conventional bidirectional splitter (f).}
\end{figure} 
It should be noted that when the air layer thickness (\(h\)) is sufficiently large, the splitter may also support guiding modes based on the total internal reﬂection (TIR) mechanism. These TIR modes are usually bidirectional and may affect the USMP bandwidth, thus, they need to be suppressed in the splitter. Their dispersion relation can be obtained by substituting $\alpha_r$ with $-ip$ in Eq.~\eqref{eq:eq3}, where $p = \sqrt{k_0^2 - k^2}$. The existence conditions for TIR modes can be  determined by analyzing the dispersion at $k = 0$, therefore their simplified dispersion equations  become
\begin{equation}
	\begin{aligned}
		\tan\left(k_0 h\right) + \sqrt{\dfrac{-\mu_v}{\varepsilon_m }}&= 0 & \text{(GDM)} \\
		\tan\left(\dfrac{k_0 h}{2}\right) + \sqrt{\dfrac{-\varepsilon_m}{k_0 \mu_v}} &= 0 & \text{(GDG)}
	\end{aligned}
	\label{eq:eq4}
\end{equation}
To suppress TIR modes in the bandgap, their low-frequency cutoff $\omega_f$ must exceed the upper bandgap boundary $\omega_b$. When $\omega_f = \omega_b$ in Eq.~(\ref{eq:eq4}), $\mu_v = 0$ and $k_0 h=\pi$; thus, their critical thickness ($h_c$) are the same for the GDM and GDG waveguides:  
\begin{equation}
	\label{eq:hc}
	h_c = \frac{\pi c}{\omega_0 + \omega_m}
\end{equation}
For the waveguide with $\omega_0 = 0.4\omega_m$, it is found that $h_{c} = 21.4\,\text{mm}$. When the air layer thickness \(h > h_c\), TIR modes emerge in the bandgap of both waveguides. To verify this, we calculated the dispersion curves of TIR and SMP modes in GDG and GDM waveguides for \(h = 25\,\text{mm}\), as shown in Figs.~\ref{fig4}(a) and ~\ref{fig4}(b). As expected, bidirectional TIR modes occur in the bandgap, and the USMP band is compressed by the bidirectional TIR
modes. Figure~\ref{fig4}(c) illustrates the USMP bandwidth \(\Delta\omega=\omega_f-\omega_{sp}\) as a function of $h$. It can be seen that the USMP bandwidth decreases with $h$ in both waveguides for \(h \geq h_c\). The results confirm that the emergence of the TIR mode significantly affects the USMP bandwidth of the beam splitter.

To further quantify the effect of the TIR mode on the splitter performance, we performed full-wave simulations of the splitter. Figure~\ref{fig4}(d) shows the transmission spectra with \(h_1 = 25\,\text{mm}\), \(h_2 = 15\,\text{mm}\), and \(h_3 = 12\,\text{mm}\). For the case of \(h_1 > h_c\), the USMP and bidirectional TIR modes are supported in the frequency ranges [0.9$\omega_m$, 1.29$\omega_m$] and [1.29$\omega_m$, 1.4$\omega_m$], marked by yellow and blue, respectively. Within the USMP band, the splitter exhibits broadband asymmetric splitting with transmission efficiency $\eta_{tot}$ near 1, similar to the results in Fig.~\ref{fig3}(f). However, in the TIR region, the splitter behaves like a conventional bidirectional splitter. The transmission coefficients oscillate with frequency and $\eta_{tot}$ drops significantly due to strong reflections at the splitter junction. For a clearer comparison between the unidirectional USMP-based splitter and the bidirectional TIR-based splitter, the simulated \(E\)-field results are displayed in Fig.~\ref{fig4}(e) and \ref{fig4}(f), respectively. It is seen that the excited wave in the unidirectional splitter can only propagate forward and is completely split into the two output waveguides without any backscattering [Fig.~\ref{fig4}(e)]. In contrast, the wave in the conventional bidirectional splitter is strongly reflected at the junction, with only a small amount of energy being split into the two output waveguides [Fig.~\ref{fig4}(f)]. Therefore, the air layer thickness of the splitter should be set to \(h_1 \leq h_c\) to suppress TIR mode in the bandgap. These results further confirm the advantage of our USMP-based splitter in achieving broadband and high-transmission-efficiency splitting, overcoming the back-reﬂection issue in conventional bidirectional splitters.

\begin{figure}[bh]
	\centering\includegraphics[width=4.2 in]{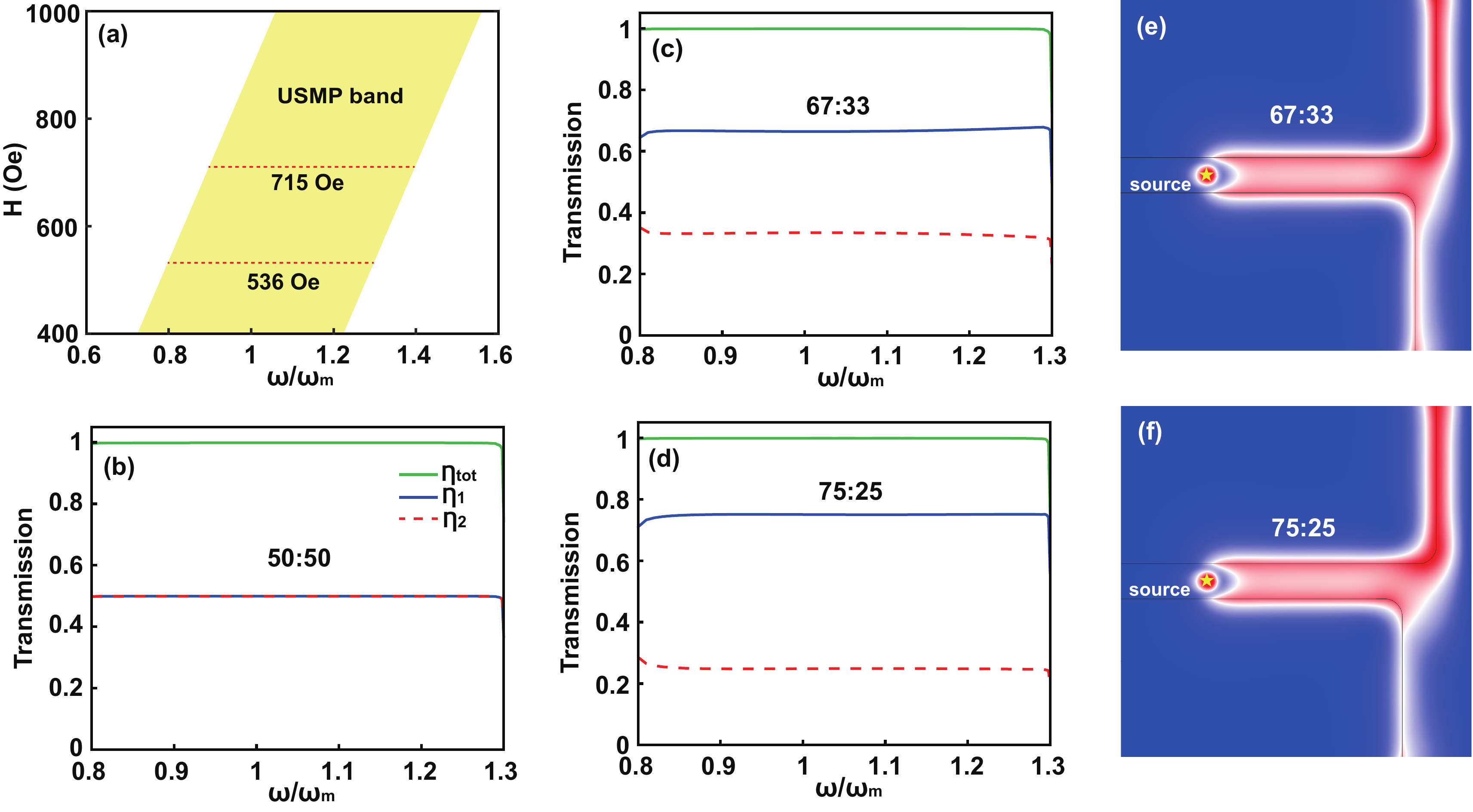}
	\caption{\label{fig5} Tunable-Band beam splitter. (a) USMP band versus magnetic field $H$ for the splitter. The red dashed lines represent $H = 715\ \mathrm{Oe}$ ($\omega_0 = 0.4\omega_m$) and $H = 536\ \mathrm{Oe}$ ($\omega_0 = 0.3\omega_m$). (b)-(d) Transmission spectra for $H = 536\ \mathrm{Oe}$. The $\eta_{1}:\eta_{2}=$ 50:50 (b), 67:33 (c), and 75:25 (d), corresponding to \(h_3 = 6\) mm, 9.5 mm and 11.8 mm, respectively. Simulated \(E\)-field distributions of the splitters with splitting ratios of 67:33 (e) and 75:25 (f), where the excitation sources (pentagram) are located at $\omega = 1.15\,\omega_m$. The other parameters identical to those in Fig.~\ref{fig3}(d).}
\end{figure}

Unlike conventional splitters with a fixed operational band, our proposed topological beam splitter achieves the tunability of the USMP band. Since the USMP band spans [$\omega_{sp}$, $\omega_{b}$], where $\omega_{sp}=0.5\omega_m + \omega_0$, $\omega_b=\omega_m + \omega_0$, and $\omega_0$ = $2\pi \gamma H$, the operational band of the splitter can be tuned by the external magnetic ﬁeld \(H\). To verify this, the relationship between USMP band and $H$ is displayed in Fig.~\ref{fig5}(a). It is seen that the USMP band shifts linearly with \(H\) from 400 to 1000 Oe. For different \(H\), broadband splitter can still be achieved. To demonstrate it, we take the variation of $H$ from \(715\)~Oe ($\omega_0 = 0.4\omega_m$) to \( 536\)~Oe ($\omega_0 = 0.3\omega_m$)  as an example to investigate the splitting characteristics. The corresponding USMP band changes from [$0.9\omega_m, 1.4\omega_m$] to [$0.8\omega_m, 1.3\omega_m$], marked by the red dashed lines. Figures~\ref{fig5}(b)–(d) show the transmission spectra for three different splitting ratios: $\eta_{1}:\eta_{2}=$ 50:50, 67:33 and 75:25, corresponding to \(h_3 = 6\) mm, 9.5 mm and 11.8 mm, respectively. Similar to the result of the 50:50 case in Fig.~\ref{fig2}(d), the bandwidth is the whole USMP band. For the 67:33 and 75:25 cases, the ranges are [$0.81\omega_m, 1.29\omega_m$] for both, covering 96\% of the USMP bandwidth, respectively. To clearly illustrate the asymmetric splitting characteristics of the beam splitter, the \(E\)-field distributions for the splitting ratios of 67:33 and 75:25 are shown in Figs.~\ref{fig5}(e) and ~\ref{fig5}(f), respectively. As expected, asymmetric splitting of electromagnetic waves at the splitter junction is achieved. These results confirm that a broadband splitter has been demonstrated to achieve magnetically tunable band by adjusting the external magnetic field.

\section{ROBUSTNESS ANALYSIS OF SPLITTER}

\begin{figure}[h]
	\centering\includegraphics[width=2.8 in]{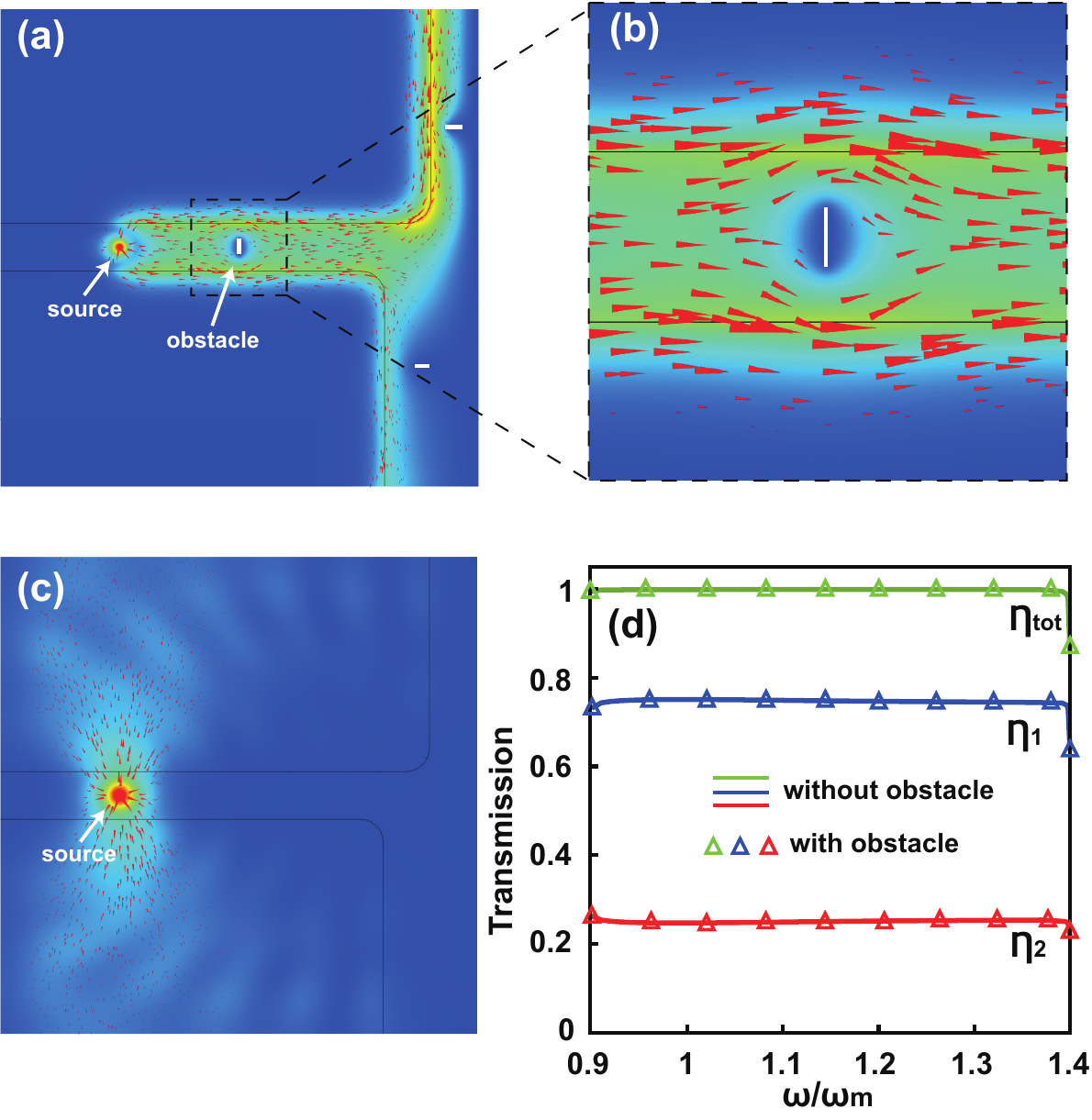}
	\caption{\label{fig6} Robust beam splitter without backscattering. (a) Simulated \(E\) field and Poynting vector distributions of the splitter for \(H = 715\) Oe, and (b) corresponding magnified view.  The white rectangles represent the PEC obstacles. (c) Simulated \(E\)-field and Poynting vector distributions of the splitter without external magnetic field (\(H = 0\) Oe). (d) Transmission spectra with (triangles) and without (solid lines) PEC obstacles. The working frequency is \( \omega = 1.15\omega_m \), and the other parameters are identical to those in Fig.~\ref{fig3}(d)}
\end{figure}

Owing to the topological property of the USMP mode, the proposed broadband splitter exhibits strong robustness against defects. To verify this, three perfect electric conductors (PEC) obstacles are introduced into the splitter [Fig.~\ref{fig3}(d)], placed in the middle of the air layers of the input and output waveguides. Figure~\ref{fig6}(a) shows the simulated \(E\)-field and Poynting vector distributions at $\omega = 1.15\omega_m$, with the PEC obstacles marked by white rectangles. Clearly, the energy fluxes bypass the PEC obstacles [Fig.~\ref{fig6}(b)] and continue to propagate unidirectionally without backscattering, indicating its strong robustness. To further quantify this, we compare the transmission spectra of the splitter with (triangles) and without (solid lines) PEC obstacles, as shown in Fig.~\ref{fig6}(d). As expected, the transmission results with and without PEC are in perfect agreement. It should be noted that when the external magnetic field is removed (\(H = 0\) Oe) in Fig.~\ref{fig6}(c), the energy fluxes are scattered into the YIG bulk because the USMP is no longer supported in the splitter. These results demonstrate the strong robustness of our splitter based on topological USMP without backscattering.

\section{DISCUSSION}
\begin{figure}[h]
	\centering\includegraphics[width=4.2 in]{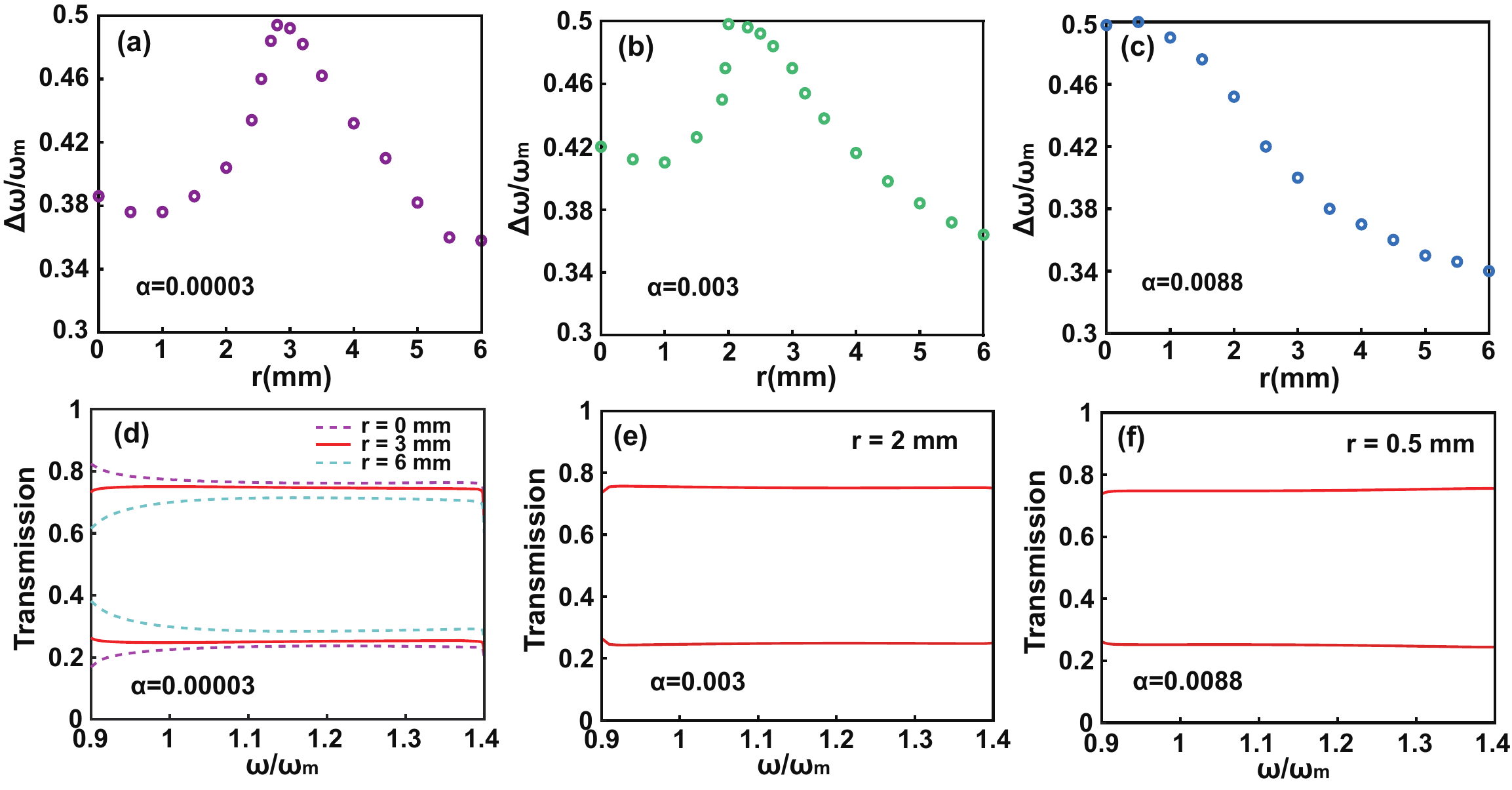}
	\caption{\label{fig7} The effect of the fillet radius \(r\) and damping coefficient $\alpha$ on the splitter. (a)-(c) Bandwidth \(\Delta \omega\) as a function of fillet radius \(r\) for different damping coefficient $\alpha$; (d)-(f)  Transmission spectra for $\alpha=0.00003$ (d), $\alpha=0.003$ (e) and $\alpha=0.0088$ (f),  respectively. The other parameters are same to those in Fig.~\ref{fig3}(d).}
\end{figure}

As demonstrated above, a broadband robust tunable splitter is realized, and its performance depends on the waveguide parameters, such as the fillet radius ($r$). Here, we further analyze the impact of the fillet radius on the splitter bandwidth.  Figure~\ref{fig7}(a) shows the bandwidth \(\Delta \omega\) as a function of $r$, with all other parameters identical to those used in Fig.~\ref{fig3}(d). It is observed that the bandwidth reaches a maximum at $r\approx 3\ \text{mm}$. To clearly show the effect of fillet radius on transmission characteristics, the transmission spectra of the splitter for $r = 0~\mathrm{mm}$, $3~\mathrm{mm}$, and $6~\mathrm{mm}$ are displaced in Fig.~\ref{fig7}(d), respectively. This result further confirms that the proposed splitter can achieve a bandwidth covering nearly the entire USMP band by setting an appropriate fillet radius.

 In addition, the  gyromagnetic materials employed in different experiments exhibit various damping coefficients \cite{liuTopologicalChernVectors2022b,liUnidirectionalGuidedwavedrivenMetasurfaces2024,wang_Topological_2025}, thus we further investigate the effect of the damping coefficients on splitter performance. As an example, two typical values are selected: $\alpha=0.003$ \cite{liUnidirectionalGuidedwavedrivenMetasurfaces2024} and $0.0088$\cite{liuTopologicalChernVectors2022b}. Figures~\ref{fig7}(b) and (c) show the splitter bandwidth $\Delta\omega$ as a function of $r$ for these two values. It is seen that the bandwidth reaches its maximum of $98\%$ at $r=2\ \text{mm}$ and $100\%$ at $r=0.5\ \text{mm}$, respectively. The corresponding transmission spectra are shown in Figs.~\ref{fig7}(e) and (f). The result demonstrated that broadband tunable splitter covering nearly 100\% of the USMP band can be achieved for different damping coefficients.

\section{CONCLUSIONS}
In summary, we have proposed a broadband and tunable beam splitter composed of GDM and GDG waveguides based on topological USMP. Unlike existing MMI-based topological beam splitters with narrowband and frequency-dependent splitting ratios, our proposed beam splitter can achieve a broadband and frequency-independent splitting ratio within a 1\% deviation across nearly the entire USMP band. Moreover, the operating band of the beam splitter can be tuned by adjusting the external magnetic field while maintaining its broadband and robust performance. Notably, the beam splitter exhibits strong robustness against obstacles and achieves nearly 100\% transmission efficiency without backscattering. These findings may have important applications in integrated photonics and reconfigurable microwave systems.

\section*{APPENDIX A: DERIVATION OF DISPERSION FORMULA}

Here, we present an analytical derivation of the SMP dispersion relation [Eq.~(\ref{eq:eq3}b)] using the GDG waveguide as an example. Similarly, the dispersion relation of SMP in the GDM waveguide can be obtained through an analogous method. This waveguide supports the TE mode ($E_z$, $H_x$, $H_y$), and the nonzero component $E_{z}$ is expressed as 
\begin{equation}
	\bm{E_z} = 
	\begin{cases}
		A_1 e^{\alpha_m y} e^{i(kx - \omega t)}, & y \geq h \\[6pt]
		(B_1 e^{-\alpha_r y} + B_2 e^{\alpha_r y}) e^{i(kx - \omega t)}, & 0 < y < h \\[6pt]
		A_2 e^{-\alpha_m y} e^{i(kx - \omega t)}, & y \leq 0
	\end{cases}
	\label{5}
\end{equation}

By solving the Maxwell's equations: $\nabla \times \bm{E} = i \omega \mu_0 \bm{\mu}^\pm_m \bm{H}$ and $\nabla \times \bm{H} = -i\omega \varepsilon \varepsilon_0 \bm{E}$, the magnetic field component $H_x$ can be obtained from $E_z$ as follow
\begin{equation}
	H_{x} = 
	\begin{cases}
		\frac{1}{i\omega\mu_0\mu_v} (\alpha_m + ik\frac{\mu_2}{\mu_1})A_1 e^{\alpha_m y} e^{i(kx - \omega t)}, & y \geq h \\
		\frac{i\alpha_r}{\omega\mu_0} \left( B_1 e^{-\alpha_r y} - B_2 e^{\alpha_r y} \right) e^{i(kx - \omega t)}, & 0 < y < h \\
		\frac{1}{i\omega\mu_0\mu_v} ( ik \frac{\mu_2}{\mu_1} - \alpha_m )A_2 e^{-\alpha_m y} e^{i(kx - \omega t)}, & y \leq 0
	\end{cases}
	\label{6}
\end{equation}

Based on the field boundary conditions, \( E_z \)  and \( H_x \)  remain continuous at the boundaries \( y =  0 \) and \( y = h \). For the continuous boundary conditions of \( E_z \) in Eq.~(\ref{5}), we obtain :
\begin{equation}
	\begin{gathered}
		A_1  = B_1 e^{-2\alpha_m h} + B_2 \\
		A_2  = B_1  + B_2 
	\end{gathered}
	\label{7}
\end{equation}

Considering the continuous boundary conditions of \( H_x \), we obtain from Eq.~(\ref{6}):  
\begin{equation}
	\begin{aligned}
		A_1 e^{\alpha_m h} \left( \alpha_m + ik \frac{\mu_2}{\mu_1} \right) &= \alpha_r \mu_v \left( B_2 e^{\alpha_r h} - B_1 e^{-\alpha_r h} \right), \\
		A_2 \left( ik \frac{\mu_2}{\mu_1} - \alpha_m \right) &= \alpha_r \mu_v \left( B_2 - B_1 \right).
	\end{aligned}
	\label{8}
\end{equation}

Thus, the dispersion relation of the SMP [Eq.~(\ref{eq:eq3}b)] can be derived by removing the coefficients \( A_1, A_2, B_1, B_2\) from Eqs.~(\ref{7}) and (\ref{8})
\begin{equation}
	e^{2\alpha_r h} - \left( \dfrac{\alpha_r \mu_v - k\dfrac{\mu_2}{\mu_1} - \alpha_m}{\alpha_r \mu_v + k\dfrac{\mu_2}{\mu_1} + \alpha_m} \right)^2 = 0
	\label{9}
\end{equation}

\begin{backmatter}
\bmsection{Funding} National Natural Science Foundation of China (12464057,62375118, 62361166627); Natural Science Foundation of Jiangxi Province(20224BAB211015, 20242BAB25039); Guangdong Basic and Applied Basic Research Foundation (2024A1515012770); Shenzhen Science and Technology Innovation Commission (20220815111105001, 202308073000209); High level of special funds (G03034K004).

\bmsection{Disclosures}
The authors declare no conflicts of interest

\bmsection{Data Availability}
Data underlying the results presented in this paper are not publicly available at this time but may be obtained from the authors upon reasonable request. The data that support the findings of this study are available from the corresponding authors upon reasonable request.
\end{backmatter}

\bibliography{reference.bib}






\end{document}